\documentclass{kluwer}

\usepackage{graphicx}

\begin{document}

\begin{article}
\begin{opening}
\title{Illuminating Protogalaxies? The Discovery of Extended Lyman-$\alpha$
Emission around a QSO at $z=4.5$}

\author{Andrew \surname{Bunker}\thanks{bunker@ast.cam.ac.uk}}
\author{Joanna \surname{Smith}}
\institute{Institute of Astronomy, University of Cambridge, Cambridge, CB3 0HA, UK}
\author{Hyron \surname{Spinrad}}
\institute{Astronomy Department,
 University of California, Berkeley, USA}
\author{Daniel \surname{Stern}}
\institute{Jet Propulsion Laboratory / California Institute of
Technology, Pasadena, USA}
\author{Stephen \surname{Warren}}
\institute{Department of Physics, Imperial College, London, UK}

\runningauthor{A.\ Bunker {\em et al.}}
\runningtitle{Extended Lyman-$\alpha$
Emission around a High-$z$ QSO}

\date{September 30, 2002}

\begin{abstract}
We have discovered extended Lyman-$\alpha$ emission around a $z=4.5$ QSO
in a deep long-slit spectrum with Keck/LRIS at moderate spectral
resolution ($R\approx 1000$). The line emission extends 5\,arcsec beyond
the continuum of the QSO and is spatially asymmetric. This extended line
emission has a spectral extent of 1000km/s, much narrower in velocity
spread than the broad Lyman-$\alpha$ from the QSO itself and slightly
offset in redshift. No evidence of continuum is seen for the extended
emission line region, suggesting that this recombination line is powered
by reprocessed QSO Lyman continuum flux rather than by local star
formation. This phenomenon is rare in QSOs which are not radio loud, and
this is the first time it has been observed at $z>4$.  It seems likely
that the QSO is illuminating the surrounding cold gas of the host
galaxy, with the ionizing photons producing Lyman-$\alpha$
fluorescence. As suggested by Haiman \& Rees (2001), this ``fuzz''
around a distant quasar may place strong constraints on galaxy formation
and the extended distribution of cold, neutral gas.
\end{abstract}

\end{opening}

\section{Introduction}

Although it is by now well-established that QSOs are hosted by galaxies,
QSOs themselves are difficult to interpret as probes of galaxy
evolution. The evolution of the QSO and galaxy populations is expected
to be linked, but in fact they exhibit very different behaviour: the
extinction-corrected Madau-Lilly diagram of volume-averaged star
formation is flat at redshifts beyond $z=1$ (Hopkins, Connolly \& Szalay
2000 AJ 120, 2843), yet the QSO luminosity function peaks sharply at
$z=2$ (Schmidt, Schneider \& Gunn 1995 AJ, 110, 68).

A quasar phase may be a natural (but brief) evolutionary stage in the
life of all massive galaxies, and theory predicts the proto-galaxy to
be enveloped by spatially-extended cold gas of temperature
$\sim$\,10,000\,K, with a radiative cooling time shorter than the
dynamical time.  Interesting questions include: What is the physical
effect of a QSO turning on within an assembling galaxy? And is this
observable?

Recent theoretical work predicts a Lyman-$\alpha$ halo with a surface
brightness of $\sim 10^{-17}\,{\rm ergs\,s^{-1}\,cm^{-2}\,arcsec^{-2}}$
(Haiman \& Rees 2001 ApJ 556, 87). Such haloes have been seen in radio
galaxies and radio-loud QSOs ({\em e.g.}, Bremer et al.\ 1992 MNRAS 258,
3) but in these cases it is probably related to outflows. Until now,
this phenomenon has yet to be seen for quasars which are not radio loud
(see Hu \& Cowie 1987 ApJLett 317, 7). If no extended emission can be
found, this may imply that QSOs only turn on when the gas has settled
into a thin disk or the cold gas has already been consumed in star
formation.

The ionizing photons from the QSO will generate recombination line
emission from optically-thick neutral hydrogen clouds around the QSO.
There will be low surface-brightness Lyman-$\alpha$ ``fuzz'' anyway from
line cooling of gas in the halo potential, and external photoionization
by UV background
({\em e.g.}, Bunker, Marleau \& Graham 1998 AJ 116, 2086), 
but the presence of the QSO will greatly enhance this. Haiman \& Rees
(2001) predict a Lyman-$\alpha$ halo extending out to a significant
fraction of virial radius ($10-100$\,kpc), corresponding to an angular
size of $\sim 3''$. The predicted surface brightness of $\sim
10^{-17}\,{\rm ergs\,s^{-1}\,cm^{-2}\,arcsec^{-2}}$ is accessible to
large telescopes with spectroscopy or deep narrow-band imaging. Another
recent theory paper by Alam \& Miralda-Escud\'{e} (2002 ApJ 568, 576)
claims 100 times fainter surface brightness and very small extent for
the extended Lyman-$\alpha$ emission ($0.4''$). The large discrepancy in
the predictions arises from different assumptions about the clumping
factor of cold gas and central concentration of the line-emitting
region.

\section{Our Observations}

We have been undertaking an extensive study of the radio-quiet quasar
PC0953+4749 at $z=4.46$ (Schneider, Schmidt \& Gunn 1991 AJ 101, 2004)
which has 3 damped Lyman-$\alpha$ systems at $z>3$ (Figure~1).  The
long-slit spectroscopy was obtained with Keck/LRIS (Oke et al.\ 1995
PASP 107, 375) for 1~hour at spectral resolution of 300km/s. Inspection
of the 2D spectrum reveals Lyman-$\alpha$ at the QSO redshift but
extended spatially beyond the continuum of the QSO (Figure~2).  This is
the first time this phenomenon has been seen at $z>4$ in a QSO which is
not radio-loud.

\begin{figure}
\centering
\resizebox{0.68\textwidth}{!}{\includegraphics*{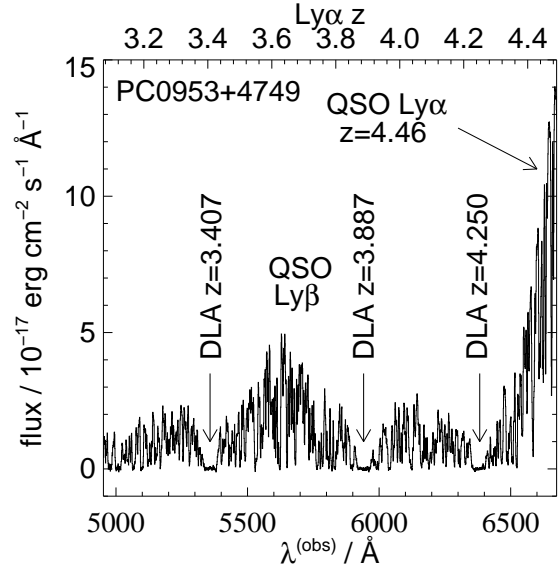}}
\caption{Our Keck/LRIS spectrum of the radio-quiet QSO  PC0953+4749 at
$z=4.46$. We confirm 3 damped Lyman-$\alpha$ systems at $z>3$.}
\end{figure}

\begin{figure}
\resizebox{0.78\textwidth}{!}{\rotatebox{270}{\includegraphics*[209,142][405,651]{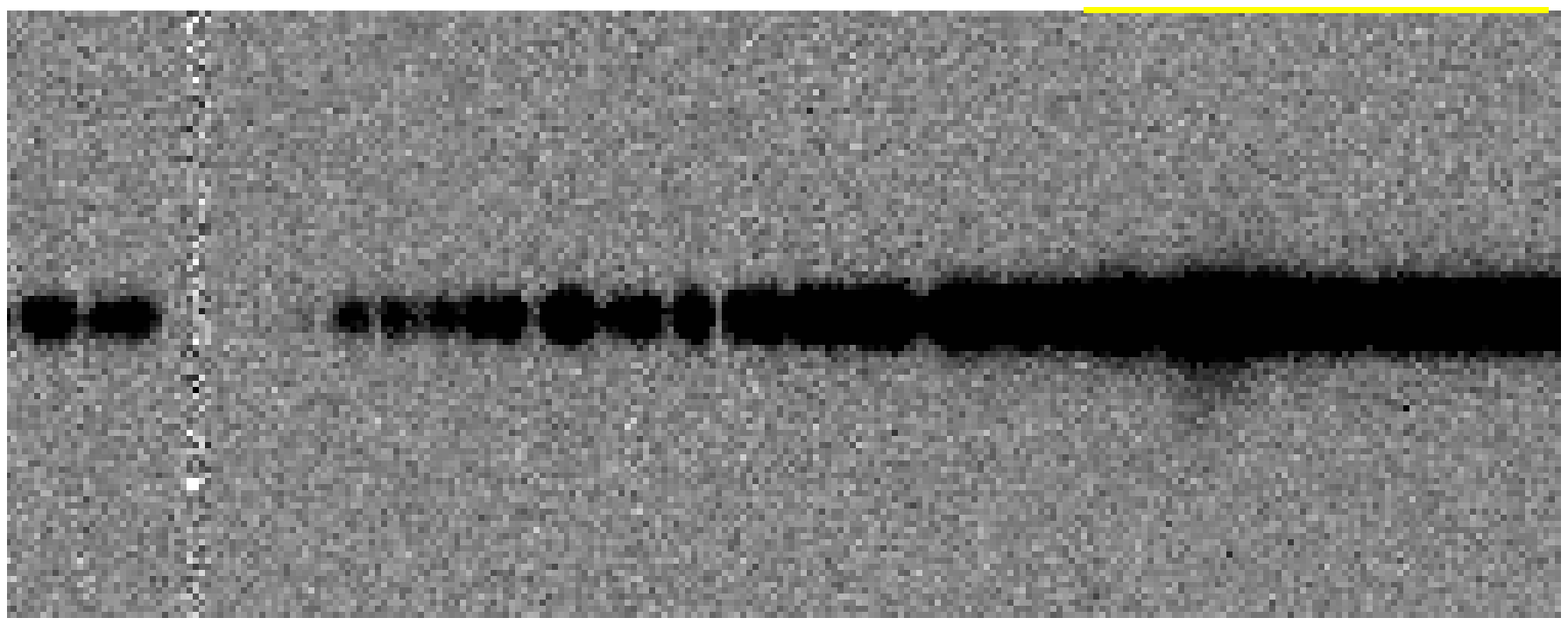}}}
\put(20,-65){\bf \vector(0,-1){30}}
\put(10,-50){\bf $20''$}
\put(20,-30){\bf \vector(0,1){30}}
\put(-230,-70){\bf $z=4.25$}
\put(-230,-40){\bf DLA}
\newline
\hspace{1cm}
\resizebox{0.78\textwidth}{!}{\rotatebox{270}{\includegraphics*[209,142][405,651]{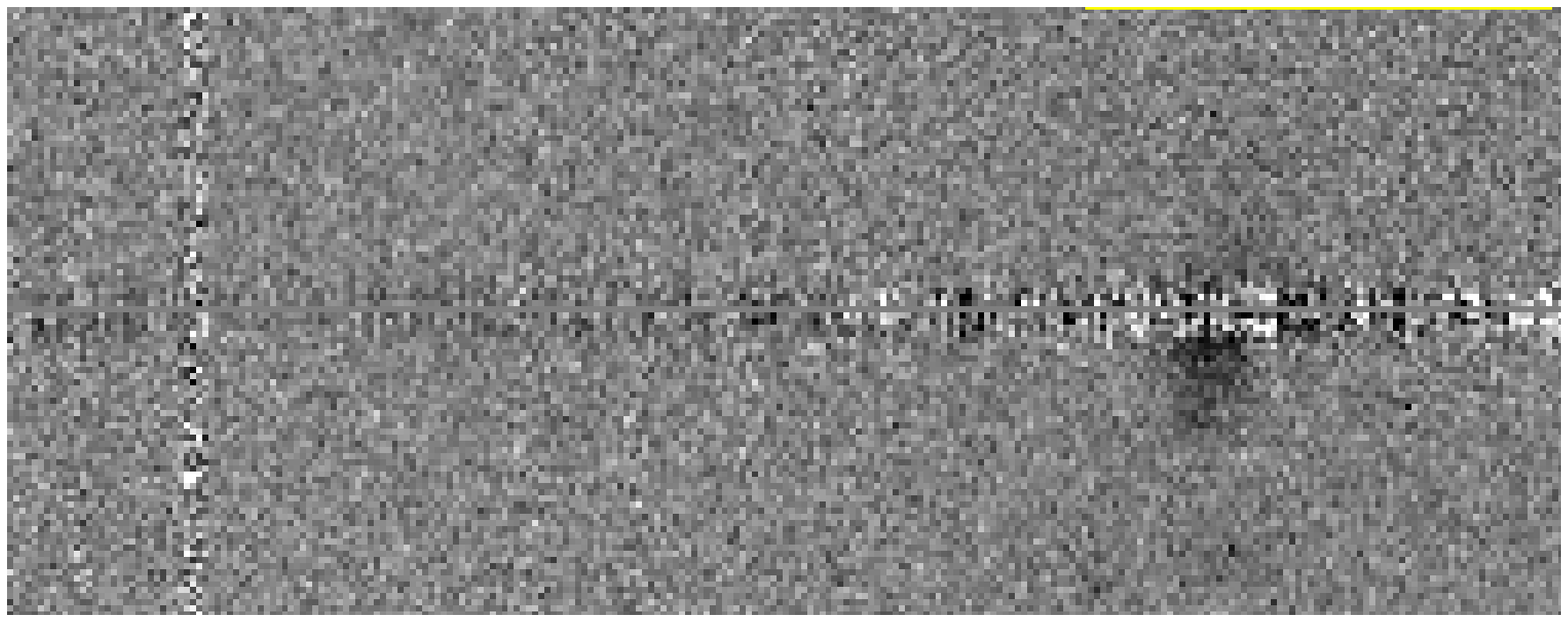}}}
\put(-150,8){\bf \vector(-1,0){100}}
\put(-143,5){\bf $400$\,\AA}
\put(-110,8){\bf \vector(1,0){100}}
\put(5,-42){\bf \vector(0,-1){32}}
\put(15,-60){\bf $5''$}
\put(5,-72){\bf \vector(0,1){32}}
\put(-70,-25){\bf Ly-$\alpha$}
\caption{The upper panel shows the 2D longslit Keck/LRIS spectrum of the
QSO PC0953+4749. Wavelength increases left-to-right, and the slit axis
is vertical. The region around Lyman-$\alpha$ from the QSO at $z=4.46$
is shown, and the damped Lyman-$\alpha$ absorption system at $z=4.25$
can also be seen near the 6363\,\AA\ sky line. The spectrum of the QSO
point source has been subtracted in the lower panel to reveal residual
Lyman-$\alpha$ emission at 6670\,\AA\ ($z=4.49$) extended over $\approx
5$\,arcsec.}
\end{figure}

This line emission extends over $\sim 5''$ beyond the QSO point spread
function. The emission is asymmetric, which implies either that gas is
clumpy, or that the radiation is beamed anisotropically.

The extended line emission (the dashed line in Figure~3) covers a
spectral extent of $\sim 1000$\,km/s FWHM. This is not a good measure of
the velocity dispersion of the gas, as this line is resonantly
broadened. The spatially extended line emission is much narrower than
Lyman-$\alpha$ from the QSO (solid line).  No evidence of continuum is
seen for the extended emission line region. This indicates that the
recombination line is probably powered by reprocessed QSO UV flux rather
than by local star formation.

\begin{figure}
\centering
\resizebox{0.78\textwidth}{!}{\includegraphics*[85,254][490,537]{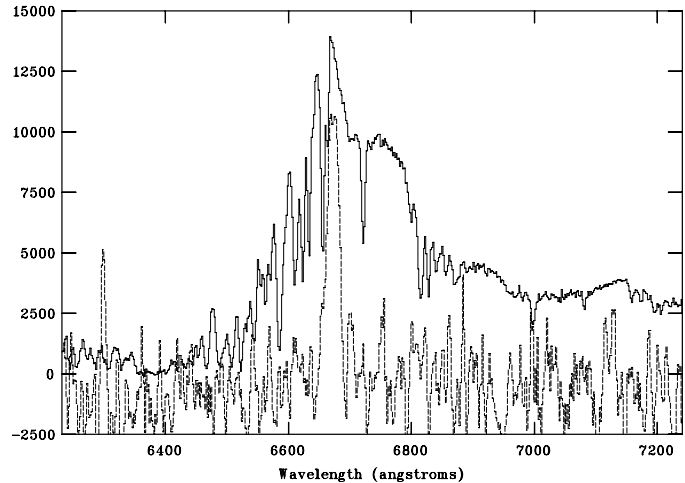}}
\caption{The extracted 1D spectrum of the QSO (solid line) compared with
that of the spatially-extended Lyman-$\alpha$ emission (dashed line,
flux scaled up by $\times$\,30).}
\end{figure}

The H\,{\scriptsize I} cloud of this host galaxy is
$>35\,h^{-1}_{70}\,kpc$ ($\Omega=0.3$). The size and surface
brightness agree more closely with the theoretical prediction of
Haiman \& Rees (2001) than with that of Alam \& Miralda-Escud\'{e}
(2002). However, we stress that this is only one example: other deep
longslit spectra of high-redshift QSOs need to be studied to see if
this extended emission is a generic feature of QSOs in young
galaxies.

\vspace{0.3cm}

{\bf Acknowledgements} AB acknowledges support from Institute of
Astronomy (Cambridge), from UC Berkeley, and from a NICMOS Fellowship. We
thank the staff of Keck Observatory.  We are grateful to Zolt\'{a}n Haiman,
Martin Haehnelt, Palle M\o ller, Martin Rees and Steve Rawlings for
illuminating discussions. We thank Shri Kulkarni, Josh Bloom, Arjun Dey
and Steve Dawson for assistance in obtaining some of the observations.

\end{article}

\end{document}